\documentclass[aps,prl,reprint,groupedaddress,showpacs,amsmath,amssymb,twocolumn,floatfix]{revtex4-1}
\usepackage{graphicx}
\usepackage{bm}
\usepackage{color}

\newcommand{\red}{\textcolor{black}}

\begin{document}

\title{Orientation of the Angular Momentum in Superfluid $^3${H}e-{A} in a Stretched  Aerogel}

\author{J.I.A. Li}
\email[]{jiali2015@u.northwestern.edu}
\author{A.M. Zimmerman}
\author{J. Pollanen}
\author{C.A. Collett}
\author{W.J. Gannon}
\author{W.P. Halperin}
\email[]{w-halperin@northwestern.edu}
\affiliation{Northwestern University, Evanston, IL 60208, USA}

\date{\today}

\begin{abstract}


Superfluid $^3${H}e-{A} in a fully characterized stretched aerogel, used in previous \red{work,}~\cite{Pol.12a} has been studied for parallel and perpendicular orientations of the magnetic field relative to the anisotropy axis of the aerogel.  Consistently, we find that an equal spin pairing state (ESP) is stabilized down to the lowest temperature. From our pulsed NMR frequency shifts as a function of temperature and tip angle, the orientation of the orbital angular momentum $\hat{l}$ has been  determined.  The aerogel anisotropy introduced by uniaxial stretching tends to align $\hat{l}$ in the axial state parallel to the strain axis, confirming the theory proposed by \red{Sauls}\cite{Sau.13} and contrary to the prediction of Volovik~\cite{Vol.08} based on an impurity calculation of Rainer and Vuorio.~\cite{Rai.77}

\end{abstract}

\maketitle

\section{Introduction}

  Since the discovery of \red{superfluid} $^3$He confined to high porosity silica aerogel,~\cite{Por.95,Spr.95} two different theories have been proposed to account for the orienting effect of aerogel anisotropy on the orbital angular momentum $\hat{l}$ in the axial state.~\cite{Vol.08,Sau.13} Specifically, for anisotropy that is introduced by uniaxial stretching,  $\hat{l}$ is predicted to be either perpendicular~\cite{Vol.08} or parallel to the strain axis.~\cite{Sau.13} For simplicity we will refer to these two theories as the easy-plane and easy-axis models respectively. The easy-plane theory assumes that aerogel strands are small cylinder-like objects with random orientation, so that the theory of Rainer and \red{Vuorio}~\cite{Rai.77} for small objects can be applied, where $\hat{l}$ is perpendicular to the axis of a single cylinder.  The orienting effect of an assembly of cylinders with a preferred orientation of  axes, possibly  induced by strain, would then  require that $\hat{l}$ be perpendicular to the strain axis as would be the case for a uniaxially stretched aerogel.  On the other hand, the easy-axis theory argues from the  point of view of symmetry.  In a uniaxial environment the choice of orbital order parameter representations is reduced from the three dimensional rotation group, SO(3),  for pure $^3$He to a combination of  two and one dimensional representations.  When uniaxial strain is present, $\hat{l}$ must be parallel to the strain.  Despite the stark difference between the two models, until now there is no clear experimental justification for either of them for aerogel with known anisotropy, either stretched or compressed.

Pulsed NMR is a powerful technique for studying the orientation of the orbital angular momentum in the axial state. The shift in the NMR spectrum, $\Delta\omega$, as a function of the angle $\beta$ by which the nuclear magnetization is tipped away from the external magnetic field, depends sensitively on the relative orientation between $\hat{l}$ and $H$ as shown in Eq.~1 and 2:

\begin{eqnarray}
    \Delta\omega_{\hat{l}\parallel H} &=& -\frac{\Omega_A^2}{2\omega_L}\cos\beta, \label{1}\\
		\Delta\omega_{\hat{l}\perp H} &=& \frac{\Omega_A^2}{2\omega_L} \frac{\left(3\cos\beta+1 \right)}{4}. \label{2}
\end{eqnarray}
where $\omega_L=\gamma H$ is the Lamor frequency, proportional to the external magnetic field $H$, and $\Omega_A$ is the longitudinal resonance frequency of the axial state. Since $\Omega_A$ is directly related to the amplitude of the order parameter $\Delta$, and goes to zero at the superfluid transition temperature $T_c$, $\Delta\omega$ in both Eq.~1 and 2 goes to zero at $T_c$. Obviously for small $\beta$, $\hat{l}\parallel H$ results in a  maximum negative frequency shift while for $\hat{l}\perp H$, a maximum positive frequency shift will be observed.

Uniform anisotropy was introduced into the aerogel sample by uniaxial stretching of $14\%$ during growth, and in this aerogel sample, an equal spin pairing (ESP) state is stabilized down to the lowest temperature.~\cite{Pol.12a} At low temperature, a new chiral-state was observed with complicated $\hat{l}$-texture. On warming from low temperature, the ESP state directly below $T_c$ was identified as the axial state from tip angle measurements of the frequency shift, Fig.~1, which match well with Eq.~2, black solid curve, indicating that the axial state is in the dipole-locked configuration with $\hat{l} \perp H$. 
The frequency shift for  small tip angle, $\beta=8^{\circ}$, as a function of temperature is shown as the blue circles in Fig.~2.  Since the strain axis is perpendicular to the field, \red{it is expected from both theoretical models that $\hat{l} \perp H$ } and  our measurements cannot discriminate between them. To definitively determine the relative orientation of $\hat{l}$ and the strain axis, we have oriented the strain axis to be parallel to the magnetic field and the results are shown in Fig. 2. and 3. In this geometry, according to the easy-plane theory, a minimum dipole energy is possible with $\hat{l} \perp H$, thus we expect a maximum positive frequency shift \red{in this case; however,}  the easy-axis theory requires $\hat{l} \parallel H$ which would lead to a maximum negative frequency shift. Our results are consistent with the latter.

\begin{figure}
\begin{center}
\includegraphics[%
  width=1\linewidth,
  keepaspectratio]{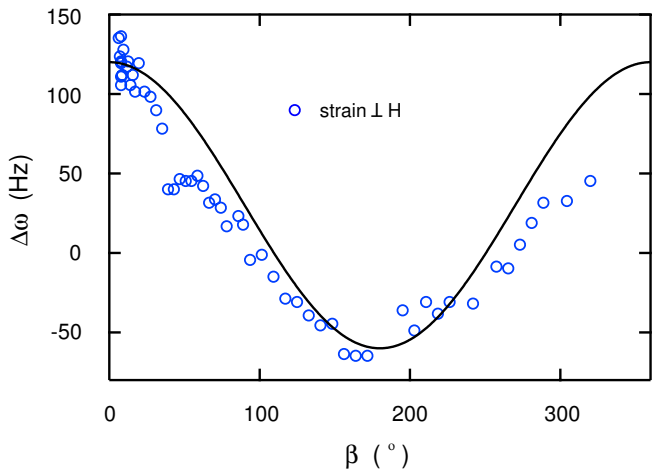}
\end{center}
\caption{(Color online) Frequency shift as a function of tip angle, $\beta$, at $\frac{T}{T_c} \approx 0.86$ with the strain axis perpendicular to the magnetic field. The black solid curve is the expected behavior for either easy-plane or easy-axis models, \red{and consequently} for this orientation of the magnetic field it is not possible to distinguish between them. }
\label{fig1}
\end{figure}

Efforts to study the effect of aerogel anisotropy with NMR frequency shifts have been reported previously.~\cite{Elb.08,Dmi.10}  However,  \red{interpretation of their} results is complicated by either an inhomogeneous distribution of  aerogel \red{strain}~\cite{Elb.08} indicated by an increase in linewidth in the superfluid state, or ambiguity in the type of anisotropy that is present in the aerogel samples owing to  lack of characterization of the aerogel.~\cite{Dmi.10} 

\section{Experimental Details}

\red{The aerogel sample in the present work is the same as that used previously}~\cite{Pol.12a} to study the orientation of $\hat{l} \perp H$.  After those experiments, we extracted the sample from the epoxy sample cell and repeated a full characterization  with optical birefringence which showed it to be unchanged and to be uniformly anisotropic with the type of anisotropy consistent with uniaxial stretching.~\cite{Pol.08,Zim.13}  We oriented the strain axis of the sample to be parallel to the external magnetic field, so that we might definitively determine the orientation of $\hat{l}$ in the axial state.  Pulsed NMR measurements were performed in the superfluid and normal fluid at a pressure $P = 26.3$ bar in a magnetic field $H = 49.9$ mT. The RF pulse generated a high homogeneity  $H_1$ field  perpendicular to the strain axis, tipping the nuclear magnetization by an angle $\beta$ away from the external field. An absorption spectrum was obtained by Fourier transformation of the free induction decay (FID) signal.  The frequency shift for all $\beta$, Fig.~3, did not depend on which part the the FID was used for calculating the Fourier transform.  The frequency shift was determined from the position of the superfluid state spectrum relative to the normal state.  The sample was cooled down to the lowest temperature, $T \approx 0.7$ mK, by adiabatic demagnetization of PrNi$_5$, NMR measurements were performed as the sample was warmed up slowly through all the superfluid transitions with a constant tip angle $\beta$, or alternatively, held at a constant temperature while varying $\beta$. Thermometry was based on a combination of $^{195}$Pt NMR, calibrated relative to the known transition temperatures of pure $^3$He \red{determined from} melting curve thermometry. 

\begin{figure}
\begin{center}
\includegraphics[%
  width=1\linewidth,
  keepaspectratio]{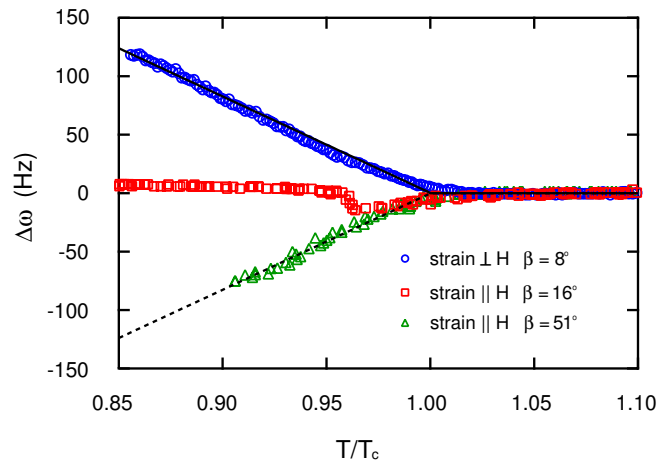}
\end{center}
\caption{(Color online) Frequency shift as a function of temperature for different aerogel orientations. For strain axis perpendicular to the magnetic field, \red{the} frequency shift with $\beta=8^{\circ}$ is shown as the blue circles. For strain axis parallel to the magnetic field, the frequency shift with $\beta=16^{\circ}$ and $51^{\circ}$ are shown as  red squares and green triangles respectively. The frequency shift with $\beta=51^{\circ}$ (green triangles) was scaled to the $\beta=0$ limit, using Eq.~1, to better compare with other data. The black solid (black dashed) \red{curves indicate} the maximum positive (negative) frequency shift \red{in the} small tip angle limit, Eq.~2 (Eq.~1).}
\label{fig2}
\end{figure}

\begin{figure}
\begin{center}
\includegraphics[%
  width=1\linewidth,
  keepaspectratio]{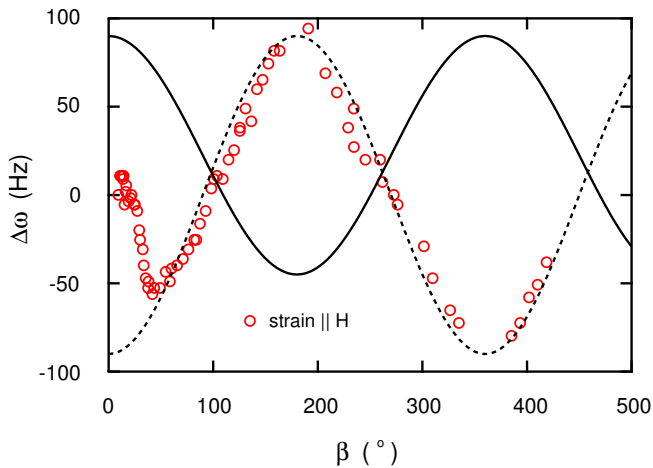}
\end{center}
\caption{(Color online) Frequency shift as a function of tip angle, $\beta$, at $\frac{T}{T_c} \approx 0.89$ with the strain axis parallel to the magnetic field. The black solid curve is the expected behavior for the easy-plane model, which does not agree with the data; while the black dashed line is the prediction of the easy-axis model, matching the data extremely well for $\beta > 50^{\circ}$.}
\label{fig3}
\end{figure}

\section{Results}

The frequency shifts as a function of reduced temperature  are shown in Fig.~2. \red{Results} from our previous experiment with the aerogel strain axis perpendicular to the magnetic field (blue circles), \red{indicate a well-defined} transition into a superfluid state.  We identified this  to be the axial state from  tip angle dependence of the frequency shift.~\cite{Pol.12a} The maximum frequency shift (black solid \red{curve}) shows \red{the} orbital angular momentum $\hat{l} \perp H$, expected from both theoretical models when the strain axis is perpendicular to the magnetic field in agreement with the data.~\cite{Vol.08,Sau.13}    

\red{With the strain axis} parallel to the magnetic field, \red{the} transition temperature  precisely \red{reproduces} that of our previous measurements.~\cite{Pol.12a}  For small $\beta$, a maximum positive frequency shift is expected from the easy-plane model,~\cite{Vol.08} while a maximum negative frequency shift is anticipated by the easy-axis model.~\cite{Sau.13}  \red{Our small} tip angle ($\beta=16^{\circ}$) measurements with strain axis parallel to the magnetic field, result in a frequency shift shown as red squares taken during a warming experiment. While the frequency shift is close to zero for most of the temperature range, disagreeing with both theoretical models, near $T_c$ a small jump to negative frequency shift is observed. For larger tip angle, \emph{e.g.}, $\beta = 51^{\circ}$, the frequency shift on warming is always negative shown as a function of temperature in Fig.~2. \red{In this figure the} frequency shift with $\beta = 51^{\circ}$ was scaled to the $\beta = 0$ limit, using Eq.~1, and matches well with the maximum negative frequency shift (black dashed \red{curve in Fig.~2}) predicted by the easy-axis model.

\red{Tip angle measurements were performed at $\frac{T}{T_c} \approx 0.89$ shown} in Fig.~3. The black solid curve is the predicted behavior when the orbital angular momentum is in the easy plane, Eq.~1\red{.  The} black dashed \red{curve} is for the easy-axis model where the orbital angular momentum is parallel to the magnetic field, according to Eq.~2.  Despite the deviation for $\beta < 50^{\circ}$, the data matches the easy-axis theory \red{extremely well} (black dashed \red{curve}), up to $\beta \approx 450^{\circ}$.  \red{The} tip angle dependent frequency shift is reproducible for both increasing and decreasing $\beta$.  \red{Consequently, we infer that the small tip angle results}  observed for $\beta < 50^{\circ}$  cannot be construed as evidence for a metastable state.  \red{For all of these measurements, the linewidth of the superfluid state}, calculated from the square root of the second moment of the spectrum, shows no increase compared to the normal \red{state.} This indicates that the orbital angular momentum is \red{uniformly oriented}  throughout the entire sample.

\section{Conclusions}
We oriented the strain axis of a stretched aerogel sample \red{taken from} our previous work~\cite{Pol.12a} to be parallel to the magnetic field in order to determine the orientation of \red{the orbital angular momentum,} $\hat{l}$.  We performed pulsed  NMR \red{measurements of} the frequency shift as a function of tip angle and temperature.  \red{Close} to $T_c$, and for  all $\beta > 50^{\circ}$,  the orbital angular momentum is parallel to the strain axis, consistent with the easy-axis theory of \red{Sauls}.~\cite{Sau.13} For $\beta < 50^{\circ}$ \red{sufficiently far} from $T_c$, the frequency shift deviates from both theoretical models. A possible mechanism for this behavior might be the competition between the dipole energy, which is minimized when $\hat{l} \perp H$, and the aerogel anisotropy that favors $\hat{l} \parallel H$. 

\begin{acknowledgements}
We are grateful for support from the National Science Foundation, DMR-1103625.
\end{acknowledgements}


\begin{thebibliography}{10}%
\makeatletter
\providecommand \@ifxundefined [1]{%
 \@ifx{#1\undefined}
}%
\providecommand \@ifnum [1]{%
 \ifnum #1\expandafter \@firstoftwo
 \else \expandafter \@secondoftwo
 \fi
}%
\providecommand \@ifx [1]{%
 \ifx #1\expandafter \@firstoftwo
 \else \expandafter \@secondoftwo
 \fi
}%
\providecommand \natexlab [1]{#1}%
\providecommand \enquote  [1]{``#1''}%
\providecommand \bibnamefont  [1]{#1}%
\providecommand \bibfnamefont [1]{#1}%
\providecommand \citenamefont [1]{#1}%
\providecommand \href@noop [0]{\@secondoftwo}%
\providecommand \href [0]{\begingroup \@sanitize@url \@href}%
\providecommand \@href[1]{\@@startlink{#1}\@@href}%
\providecommand \@@href[1]{\endgroup#1\@@endlink}%
\providecommand \@sanitize@url [0]{\catcode `\\12\catcode `\$12\catcode
  `\&12\catcode `\#12\catcode `\^12\catcode `\_12\catcode `\%12\relax}%
\providecommand \@@startlink[1]{}%
\providecommand \@@endlink[0]{}%
\providecommand \url  [0]{\begingroup\@sanitize@url \@url }%
\providecommand \@url [1]{\endgroup\@href {#1}{\urlprefix }}%
\providecommand \urlprefix  [0]{URL }%
\providecommand \Eprint [0]{\href }%
\providecommand \doibase [0]{http://dx.doi.org/}%
\providecommand \selectlanguage [0]{\@gobble}%
\providecommand \bibinfo  [0]{\@secondoftwo}%
\providecommand \bibfield  [0]{\@secondoftwo}%
\providecommand \translation [1]{[#1]}%
\providecommand \BibitemOpen [0]{}%
\providecommand \bibitemStop [0]{}%
\providecommand \bibitemNoStop [0]{.\EOS\space}%
\providecommand \EOS [0]{\spacefactor3000\relax}%
\providecommand \BibitemShut  [1]{\csname bibitem#1\endcsname}%
\let\auto@bib@innerbib\@empty
\bibitem [{\citenamefont {Pollanen}\ \emph {et~al.}(2012)\citenamefont
  {Pollanen}, \citenamefont {Li}, \citenamefont {Collett}, \citenamefont
  {Gannon}, \citenamefont {Halperin},\ and\ \citenamefont {Sauls}}]{Pol.12a}%
  \BibitemOpen
  \bibfield  {author} {\bibinfo {author} {\bibfnamefont {J.}~\bibnamefont
  {Pollanen}}, \bibinfo {author} {\bibfnamefont {J.~I.~A.}\ \bibnamefont {Li}},
  \bibinfo {author} {\bibfnamefont {C.~A.}\ \bibnamefont {Collett}}, \bibinfo
  {author} {\bibfnamefont {W.~J.}\ \bibnamefont {Gannon}}, \bibinfo {author}
  {\bibfnamefont {W.~P.}\ \bibnamefont {Halperin}}, \ and\ \bibinfo {author}
  {\bibfnamefont {J.~A.}\ \bibnamefont {Sauls}},\ }\href@noop {} {\bibfield
  {journal} {\bibinfo  {journal} {Nature Phys.}\ }\textbf {\bibinfo {volume}
  {8}},\ \bibinfo {pages} {317} (\bibinfo {year} {2012})}\BibitemShut {NoStop}%
\bibitem [{\citenamefont {Sauls}()}]{Sau.13}%
  \BibitemOpen
  \bibfield  {author} {\bibinfo {author} {\bibfnamefont {J.~A.}\ \bibnamefont
  {Sauls}},\ }\href@noop {} {\bibinfo  {journal} {to be published}\
  }\BibitemShut {NoStop}%
\bibitem [{\citenamefont {Volovik}(2008)}]{Vol.08}%
  \BibitemOpen
\bibfield  {journal} {  }\bibfield  {author} {\bibinfo {author} {\bibfnamefont
  {G.~E.}\ \bibnamefont {Volovik}},\ }\href@noop {} {\bibfield  {journal}
  {\bibinfo  {journal} {J. Low Temp. Phys.}\ }\textbf {\bibinfo {volume}
  {150}},\ \bibinfo {pages} {453} (\bibinfo {year} {2008})}\BibitemShut
  {NoStop}%
\bibitem [{\citenamefont {Rainer}\ and\ \citenamefont {Vuorio}(1977)}]{Rai.77}%
  \BibitemOpen
  \bibfield  {author} {\bibinfo {author} {\bibfnamefont {D.}~\bibnamefont
  {Rainer}}\ and\ \bibinfo {author} {\bibfnamefont {M.}~\bibnamefont
  {Vuorio}},\ }\href@noop {} {\bibfield  {journal} {\bibinfo  {journal} {J.
  Phys. C: Solid State Phys.}\ }\textbf {\bibinfo {volume} {10}},\ \bibinfo
  {pages} {3093} (\bibinfo {year} {1977})}\BibitemShut {NoStop}%
\bibitem [{\citenamefont {Porto}\ and\ \citenamefont {Parpia}(1995)}]{Por.95}%
  \BibitemOpen
  \bibfield  {author} {\bibinfo {author} {\bibfnamefont {J.~V.}\ \bibnamefont
  {Porto}}\ and\ \bibinfo {author} {\bibfnamefont {J.~M.}\ \bibnamefont
  {Parpia}},\ }\href@noop {} {\bibfield  {journal} {\bibinfo  {journal} {Phys.
  Rev. Lett.}\ }\textbf {\bibinfo {volume} {74}},\ \bibinfo {pages} {4667}
  (\bibinfo {year} {1995})}\BibitemShut {NoStop}%
\bibitem [{\citenamefont {Sprague}\ \emph {et~al.}(1995)\citenamefont
  {Sprague}, \citenamefont {Haard}, \citenamefont {Kycia}, \citenamefont
  {Rand}, \citenamefont {Lee}, \citenamefont {Hamot},\ and\ \citenamefont
  {Halperin}}]{Spr.95}%
  \BibitemOpen
  \bibfield  {author} {\bibinfo {author} {\bibfnamefont {D.~T.}\ \bibnamefont
  {Sprague}}, \bibinfo {author} {\bibfnamefont {T.~M.}\ \bibnamefont {Haard}},
  \bibinfo {author} {\bibfnamefont {J.~B.}\ \bibnamefont {Kycia}}, \bibinfo
  {author} {\bibfnamefont {M.~R.}\ \bibnamefont {Rand}}, \bibinfo {author}
  {\bibfnamefont {Y.}~\bibnamefont {Lee}}, \bibinfo {author} {\bibfnamefont
  {P.~J.}\ \bibnamefont {Hamot}}, \ and\ \bibinfo {author} {\bibfnamefont
  {W.~P.}\ \bibnamefont {Halperin}},\ }\href@noop {} {\bibfield  {journal}
  {\bibinfo  {journal} {Phys. Rev. Lett.}\ }\textbf {\bibinfo {volume} {75}},\
  \bibinfo {pages} {661} (\bibinfo {year} {1995})}\BibitemShut {NoStop}%
\bibitem [{\citenamefont {Elbs}\ \emph {et~al.}(2008)\citenamefont {Elbs},
  \citenamefont {Bunkov}, \citenamefont {Collin},\ and\ \citenamefont
  {Godfrin}}]{Elb.08}%
  \BibitemOpen
  \bibfield  {author} {\bibinfo {author} {\bibfnamefont {J.}~\bibnamefont
  {Elbs}}, \bibinfo {author} {\bibfnamefont {Y.~M.}\ \bibnamefont {Bunkov}},
  \bibinfo {author} {\bibfnamefont {E.}~\bibnamefont {Collin}}, \ and\ \bibinfo
  {author} {\bibfnamefont {H.}~\bibnamefont {Godfrin}},\ }\href@noop {}
  {\bibfield  {journal} {\bibinfo  {journal} {Phys. Rev. Lett.}\ }\textbf
  {\bibinfo {volume} {100}},\ \bibinfo {pages} {215304} (\bibinfo {year}
  {2008})}\BibitemShut {NoStop}%
\bibitem [{\citenamefont {Dmitriev}\ \emph {et~al.}(2010)\citenamefont
  {Dmitriev}, \citenamefont {Krasnikhin}, \citenamefont {Mulders},
  \citenamefont {Senin}, \citenamefont {Volovik},\ and\ \citenamefont
  {Yudin}}]{Dmi.10}%
  \BibitemOpen
  \bibfield  {author} {\bibinfo {author} {\bibfnamefont {V.~V.}\ \bibnamefont
  {Dmitriev}}, \bibinfo {author} {\bibfnamefont {D.~A.}\ \bibnamefont
  {Krasnikhin}}, \bibinfo {author} {\bibfnamefont {N.}~\bibnamefont {Mulders}},
  \bibinfo {author} {\bibfnamefont {A.~A.}\ \bibnamefont {Senin}}, \bibinfo
  {author} {\bibfnamefont {G.~E.}\ \bibnamefont {Volovik}}, \ and\ \bibinfo
  {author} {\bibfnamefont {A.~N.}\ \bibnamefont {Yudin}},\ }\href@noop {}
  {\bibfield  {journal} {\bibinfo  {journal} {JETP Lett.}\ }\textbf {\bibinfo
  {volume} {91}},\ \bibinfo {pages} {599} (\bibinfo {year} {2010})}\BibitemShut
  {NoStop}%
\bibitem [{\citenamefont {Pollanen}\ \emph {et~al.}(2008)\citenamefont
  {Pollanen}, \citenamefont {Shirer}, \citenamefont {Blinstein}, \citenamefont
  {Davis}, \citenamefont {Choi}, \citenamefont {Lippman}, \citenamefont
  {Lurio},\ and\ \citenamefont {Halperin}}]{Pol.08}%
  \BibitemOpen
  \bibfield  {author} {\bibinfo {author} {\bibfnamefont {J.}~\bibnamefont
  {Pollanen}}, \bibinfo {author} {\bibfnamefont {K.~R.}\ \bibnamefont
  {Shirer}}, \bibinfo {author} {\bibfnamefont {S.}~\bibnamefont {Blinstein}},
  \bibinfo {author} {\bibfnamefont {J.~P.}\ \bibnamefont {Davis}}, \bibinfo
  {author} {\bibfnamefont {H.}~\bibnamefont {Choi}}, \bibinfo {author}
  {\bibfnamefont {T.~M.}\ \bibnamefont {Lippman}}, \bibinfo {author}
  {\bibfnamefont {L.~B.}\ \bibnamefont {Lurio}}, \ and\ \bibinfo {author}
  {\bibfnamefont {W.~P.}\ \bibnamefont {Halperin}},\ }\href@noop {} {\bibfield
  {journal} {\bibinfo  {journal} {J. Non-Crystalline Solids}\ }\textbf
  {\bibinfo {volume} {354}},\ \bibinfo {pages} {4668} (\bibinfo {year}
  {2008})}\BibitemShut {NoStop}%
\bibitem [{\citenamefont {Zimmerman}\ \emph {et~al.}(2013)\citenamefont
  {Zimmerman}, \citenamefont {Specht}, \citenamefont {Ginzburg}, \citenamefont
  {Pollanen}, \citenamefont {Li}, \citenamefont {Collett}, \citenamefont
  {Gannon},\ and\ \citenamefont {Halperin}}]{Zim.13}%
  \BibitemOpen
  \bibfield  {author} {\bibinfo {author} {\bibfnamefont {A.~M.}\ \bibnamefont
  {Zimmerman}}, \bibinfo {author} {\bibfnamefont {M.~G.}\ \bibnamefont
  {Specht}}, \bibinfo {author} {\bibfnamefont {D.}~\bibnamefont {Ginzburg}},
  \bibinfo {author} {\bibfnamefont {J.}~\bibnamefont {Pollanen}}, \bibinfo
  {author} {\bibfnamefont {J.~I.~A.}\ \bibnamefont {Li}}, \bibinfo {author}
  {\bibfnamefont {C.~A.}\ \bibnamefont {Collett}}, \bibinfo {author}
  {\bibfnamefont {W.~J.}\ \bibnamefont {Gannon}}, \ and\ \bibinfo {author}
  {\bibfnamefont {W.~P.}\ \bibnamefont {Halperin}},\ }\href@noop {} {\bibfield
  {journal} {\bibinfo  {journal} {J. Low Temp. Phys.}\ }\textbf {\bibinfo
  {volume} {171}},\ \bibinfo {pages} {745} (\bibinfo {year}
  {2013})}\BibitemShut {NoStop}%
\end{thebibliography}
\end{document}